\documentstyle[psfig,aps]{revtex}
\begin{document}
\draft
\title{Instability of Shear Waves in an Inhomogeneous Strongly
Coupled Dusty Plasma}
\author{Amruta Mishra\footnote[1]
{Electronic mail: am@plasma.ernet.in},
P. K. Kaw and A. Sen}
\address{Institute for Plasma Research,\\
Bhat -- 382 428, India} \maketitle
\begin{abstract}

It is demonstrated that low frequency shear modes in a strongly coupled,
inhomogeneous, dusty plasma can grow on account of an instability
involving the dynamical charge fluctuations of the dust grains. The
instability is driven by the gradient of the equilibrium dust charge
density and is associated with the finite charging time of the dust
grains. The present calculations, carried out in the generalized hydrodynamic
viscoelastic formalism, also bring out important modifications in
the threshold and growth rate of the instability due to collective
effects associated with coupling to the compressional mode.

\end{abstract}

\vspace{0.2in}

\section{Introduction}
Dusty plasmas are of great interest because of their possible applications
to a number of fields of contemporary research such as plasma
astrophysics of interplanetary and interstellar matter,
fusion research, plasmas used for semiconductor etching, arc plasmas used
to manufacture fine metal and ceramic powders, plasmas in simple flames
etc.\cite{whipple}. It is now widely recognized that
the dust component in these plasmas is often in the strongly coupled
coulomb regime with the parameter, $\Gamma \simeq (Z_de)^2/T_d d$,
typically taking values much greater than unity ($(-Z_d e)$ is the
charge on the dust particle, $d \simeq n_d^{-1/3}$ is the
interparticle distance and $T_d$ is the temperature of the dust
component). This leads to many novel physical effects such as the
formation of dust plasma crystals \cite{ikezi}, modified dispersion
of the compressional waves \cite{rosenberg,kaw}, the existence of the
transverse shear waves \cite{kaw} etc.  Many of these novel features
have now been verified by experiments and computer simulations
\cite{pieper}.

Recently, an experiment on the self--excitation of the vertical motion
of the dust particles trapped in a plasma sheath boundary, has been
reported \cite {takamura}. The physics of this excitation is related
to charging of the dust particles by the inflow of ambient plasma currents
in the inhomogeneous plasma sheath and the delay resulting because of the
finite time required by the charging process to bring the dust charge
to its ambient steady state value. In this paper, we
demonstrate that the same physical mechanism can be used for the
excitation of the transverse shear modes in an inhomogeneous strongly
 coupled dusty plasma. Using a generalized hydrodynamic viscoelastic
formalism \cite{ich} to describe the strongly coupled dusty plasma and
incorporating the novel feature of time variation of the dust
charge through a charge dynamics equation \cite{jana}, we have
derived a general dispersion relation for low frequency shear
and compressional modes in the plasma. We find that in a plasma
with finite gradients of the  equilibrium dust charge density,
the two modes are coupled and we show that the shear mode is
driven unstable if certain threshold values are exceeded.\\

Our paper is organized as follows. In the next section we briefly
discuss the equilibrium of an inhomogeneous dusty plasma that is
confined against gravity by the electric field of a plasma
sheath. In such a configuration dust particles of varying sizes
and charges arrange themselves in horizontal layers at different
heights to form a nonuniform cloud \cite {chu,nunomura}. In section 3
we carry out a linear stability analysis of such an equilibrium
in the framework of the generalized hydrodynamic equations.  The
dispersion relation of the coupled shear wave and compression
wave is solved analytically (in simple limits) as well as
numerically in section 4. The physical mechanism of the shear wave
instability is also discussed and the modifications in the
threshold and growth rate brought about by the coupling to
compressive waves are elucidated. Section 5 is devoted to a
summary and discussion of the principal results.\\

\section{Dust Cloud Equilibrium}

We consider an inhomogeneous sheath equilibrium in which the dust
particles are suspended with electric field forces balancing the
gravitational force on the particle and in which the dust charge
$(-Z_de)$ and dust size $r_d$ are both functions of the vertical
distance $z$. Then the force balance equation gives,
\begin{equation}
Z_d(z)e E_0(z) = \frac{4}{3}\pi r_d(z)^3 \rho g,
\label{frcbln}
\end{equation}
where, $\;\rho \;$, $\;g\;$, $\;E_{0}\;$ refer respectively to
the dust mass density, gravitational acceleration and the sheath
electric field.  For particle sizes of the order of a few
microns, other forces acting on the particle (such as the drag
and viscous forces) are about an order of magnitude smaller than
the electric and gravitational forces and can be neglected for
the equilibrium calculation\cite{nunomura}.  Note that for dust particles of a
uniform size (monodispersive size distribution) the above
equilibrium can only be attained at one vertical point leading to
a monolayer of dust.  A dispersion in sizes leads to a large
number of layers resulting in a nonuniform dust cloud with a
gradient in the equilibrium charge $(-Z_{d}e)$ and the dust size
$r_{d}$. The electric field $E_{0}$ is determined by the sheath
condition,
\begin{equation}
\frac{dE_{0}}{dz} = -4\pi e \left ( n_{e} - n_{i} + Z_{d}n_{d} \right )
\label{pois1}
\end{equation}
where $\;\;n_{e,i,d}\;\;$ are the local electron, ion and dust
densities respectively.
The charge $(-Z_{d}e)$ on a dust particle in the sheath region
is given by $(-Z_{d}e) = C_{d}(\phi_f- \phi) $ where $C_{d}$ is
the capacitance, $\phi_f$ is the floating potential at the
surface of the dust particle and $\phi$ is the bulk plasma
potential. For a spherical
dust particle $C_{d} = r_{d}$, and the floating potential can be
determined by the steady state condition from the dust charging
equation, namely,\cite{jana}
\begin{equation}
I_e+I_i = 0
\label{chgeqlbm}
\end{equation}
\noindent where the electron and ion currents impinging on the
dust particle are given by \cite{whipple}
\begin{mathletters}
\begin{equation}
I_e=-\pi r_{d}^2 e \bigg(\frac {8kT_e}{\pi m_e}\bigg)^{1/2} n_e
\exp\bigg [ \frac {e}{kT_e}(\phi_f-\phi)\bigg],
\end{equation}
\begin{equation}
I_i=\pi r_{d}^2 e \bigg (\frac {8kT_i}{\pi m_i}\bigg )^{1/2} n_i
\bigg [1- \frac {2e}{kT_i+ m_{i}v_{id}^{2}}(\phi_f-\phi)\bigg ].
\end{equation}
\end{mathletters}
\noindent Here $T_e$ and $T_i$ are the electron and ion
temperatures, $m_{i}$ is the ion mass and $v_{id}$ is the mean
drifting velocity of the ions in the electric field of the sheath
(it is assumed to be the ion sound velocity at the sheath edge).
We also assume that the dust particles have much smaller thermal
velocities than the electrons and ions.\\

Equations (\ref{frcbln} - \ref{chgeqlbm}) selfconsistently
determine the equilibrium of the dust cloud. Such clouds have
been experimentally observed in a number of
experiments\cite{chu,nunomura}. In \cite{nunomura} theoretical
modeling along the lines discussed above, agree very well with
the experimental observations of clouds formed with polydispersive
particle size distribution of dust particles trapped in the plasma
sheath region. A typical equilibrium variation of the dust
particle size with the vertical distance, when the Child Langmuir
law holds for the plasma sheath potential, is given
as\cite{nunomura},
\begin{equation}
r_{d} = \left (\frac{3(\phi_{f}-\phi)}{4\pi\rho g} \right )^{1/2}
\left ( \frac{6\pi en_{s}C_{s}}{\mu_{i}} \right )^{1/3}
(\delta - z )^{1/3}
\label{size}
\end{equation}
where $\;n_{s}\;$, $\;C_{s}\;$ are the plasma density and the
ion sound velocity, $\delta$ is the thickness of the sheath and
$\mu_{i} = (e\lambda_{i-n} /m_{i})^{1/2}$ with $\lambda_{i-n}$
representing the mean free path of ions colliding with the
background neutrals.  Using (\ref{size}) we can obtain the
corresponding $z$ variation for $Z_{d}$.\\

As discussed in detail in \cite{nunomura}, this dust cloud
equilibrium is confined to the plasma sheath boundary region in
the potential well created from the upward electrostatic and
downward gravitational forces. Note that the force balance
equation (\ref{frcbln}) does not prevent the particles from
oscillating about their mean positions especially if they have
significant kinetic energy or temperature. However their mean
positions are at various vertical distances and the mean $Z_{d}$
is a function of $z$. This is reminiscent of particle gyrations
in a magnetic field. If we consider wave motions in which dust
oscillation excursions are much smaller than wavelengths, we can
use a fluid theory to analyze such behaviour. In the next section,
we adopt this view point and carry out a linear stability analysis
of the equilibrium discussed above to low frequency wave perturbations.\\

\section{Linear Stability Analysis}
For low frequency perturbations in the regime
$kv_{thd}<<\omega<<kv_{the},kv_{thi}$, where $v_{thd}$, $v_{the}$
and $v_{thi}$ are the thermal velocities of the dust, electron
and ion components respectively, the electron and ion responses
obey the Boltzmann law which can be simply obtained from an
ordinary hydrodynamic representation. The dust component on the
other hand can be in the strongly coupled regime for which a
proper description is provided by the generalized viscoelastic
formalism. Using such a description a general dispersion relation
for low frequency waves (with typical wavelengths longer than any
lattice spacings) was obtained in \cite{kaw} for longitudinal
sound waves and transverse shear waves. The shear modes exist in
a strongly coupled dusty plasma because of elasticity effects
introduced by strong correlations \cite{kaw}. Our objective in
this work is to look for the effect of dust charge dynamics on
these shear modes in the strongly coupled regime.  As
demonstrated in our earlier work \cite {kaw}, the coupling of the
low frequency shear modes to transverse electromagnetic
perturbations is finite but negligibly small; we ignore this
coupling here. However, introduction of the dust charge dynamics
in the inhomogeneous plasma leads to a coupling of the low
frequency shear and compressional modes; thus the space charge
dynamics and quasineutrality condition play an important role in
describing the perturbations. The basic equations for the dust
fluid \cite{ich} we work with, are the continuity equation,
\begin{equation}
\frac {\partial}{\partial t}\delta n_d+n_{d0} \vec {\nabla} \cdot
\delta \vec{u}_{d} + \frac{n_{d0}}{M}( \delta \vec{u}_{d}
\cdot \vec {\nabla}) M  =  0,
\label{cont}
\end{equation}
the equation of motion,
\begin{eqnarray}
&&  \Big(1+\tau_m \frac {\partial}{\partial t} \Big) \Bigg[\Big(\frac
 {\partial}{\partial t}+\nu \Big)\vec {\delta u_d}
+ \frac {{\vec \bigtriangledown }\delta P}{Mn_{d0}}
+\frac {Z_d e}{M}\vec {\delta E}
 \nonumber \\
&+&\frac {\delta Z_d}{M}e \vec {E_0} \Bigg]
= \frac {1}{Mn_{d0}}\Big [\eta {\vec \bigtriangledown}^2 \vec {\delta
u_d}
+\big(\zeta+\frac {\eta}{3}\big)
{\vec \bigtriangledown} ({\vec \bigtriangledown} \cdot \vec {\delta
u_d})
\Big ],
\label {gh}
\end{eqnarray}
and the equation of state, $(\partial P/\partial n)_T\equiv MC_d^2$,
given in terms of the compressibility, $\mu_d$, as \cite{kaw}
\begin{equation}
\mu_d \equiv \frac {1}{T_d}\Big (\frac {\partial P}{\partial n}\Big) _T
=1+\frac {u(\Gamma)}{3} +\frac {\Gamma}{9}
\frac {\partial u(\Gamma)}{\partial \Gamma},
\label {eos}
\end{equation}
with the excess internal energy of the system given by
 the fitting formula \cite {slattery}
\begin{equation}
u(\Gamma)=-0.89 \Gamma+0.95 \Gamma^{1/4}+0.19 \Gamma^{-1/4}-0.81.
\end{equation}
\noindent In the above, $M$ is the dust mass, $\nu$ is the
dust--neutral collision frequency, $\delta u_d$, $\delta n_d$ and
$\delta Z_d$ are the perturbations
 in the dust velocity, number density and dust
charge, $\delta P$, $\delta E$ are the pressure and electric field
perturbations, $n_{d0}$ and $Z_d$ are the equilibrium number
density and charge for the dust and $E_0$ is the unperturbed
electric field. $\eta$ and $\zeta$ refer to the coefficients of
the shear and bulk viscosities and $\tau_m$ is the viscoelastic
relaxation time. Note that in the continuity equation we have a
contribution from the equilibrium inhomogeneity in the dust mass
distribution (arising from the size dispersion of the particles).
This term as we shall see later modifies the real
frequency of the shear waves.\\

These equations are supplemented with the dynamical equation for
the dust charge perturbations which, for perturbations
with phase velocity much smaller than the electron and ion thermal
velocities, is given as \cite{jana}
\begin{equation}
\frac {\partial}{\partial t}(\delta Z_d)+\vec {\delta u_d}\cdot {\vec
\bigtriangledown} Z_d+\eta_c \delta Z_d=-\frac{|I_{e0}|}{e}
\Bigg (\frac {\delta n_i}{n_{i0}}
-\frac{\delta n_e}{n_{e0}} \Bigg),
\label {fluct}
\end{equation}
where, $\eta_c=\Big (e|I_{e0}|/C\Big) \Big (1/{T_e}+1/{w_0}
\Big)$ is the inverse of charging time of dust grains and
$w_0=T_i-e(\phi_f-\phi)_0$. Note that the second term on the left
hand side of eq.(\ref{fluct}) arises because of the inhomogeneity
of the mean charge on the dust particles; as shall be shown
later, this is the critical term responsible for the instability.
It is also obvious that the dust charge variation in space will
lead to shielding by electrons and ions with the associated
coupling of the perturbation to dust compressional modes. We must
thus extend the above set of equations to include the
quasi-neutrality condition,
\begin{equation}
\delta n_e+Z_d \delta n_d+n_{d0} \delta Z_d-\delta n_i\simeq0,
\label{quasi}
\end{equation}
and the equation describing the electron and ion density
perturbations in terms of the potential, as
\begin{equation}
\frac {\delta n_e}{n_{e0}}=\frac {e\delta \phi}{T_e};\;\;\;\
\frac {\delta n_i}{n_{i0}}=-\frac {e\delta \phi}{T_i}.
\label{boltz}
\end{equation}
\noindent
These are the Boltzmann relations which arise
whenever the perturbations satisfy $\omega<<kv_{the},kv_{thi}$.

We shall next derive the dispersion relation for the low
frequency mode. We may note that the typical time scale for the
decay of the charge fluctuations for the dust can be very small
\cite{takamura}, with $\eta_c>>\omega$ and we shall work in that
limit.  We use the local approximation (wave lengths smaller than
characteristic equilibrium scale lengths) and choose the
propagation vector for the wave perturbation as $\vec k=(k,0,0)$,
the perturbed dust velocity, $\vec {\delta u_d}=(\delta
u_1,0,\delta u_3)$ and the perturbation in the electric field as
$\vec {\delta E}=-ik\delta \phi (1,0,0)$. Using the continuity
equation (\ref{cont}) and the equations (\ref{fluct}) --
(\ref{boltz}), and after some simple algebra, one obtains the
fluctuation in the dust charge and the potential as
\begin{mathletters}
\begin{equation}
\delta Z_d=\frac {a_1}{D}\left (\frac {k}{\omega}\right )\delta u_1
+\Big(\frac {a_2}{D}+\frac {a_3}{(i\omega)D}\Big) \delta u_3,
\end{equation}
\begin{equation}
\delta \phi=-\frac {Z_d n_{d0} \eta_c}{eD}\bigg(\frac {k}{\omega}\bigg)
\delta u_1+\frac {n_{d0}}{eD} \Big (Z_d'
-\frac {Z_d M'\eta _c}{M(i\omega)} \Big)\delta u_3,
\end{equation}
\label{delz}
\end{mathletters}
where,
\begin{eqnarray}
a_1 &=&-\frac {|I_{e0}|}{e}\Bigg (\frac {1}{T_e}+\frac {1}{T_i}\Bigg)
 Z_d n_{d0};\;
a_2=-Z_d' \Bigg (\frac {n_{e0}}{T_e}+\frac {n_{i0}}{T_i}\Bigg),\nonumber
\\
&a_3&=
-\frac {|I_{e0}|}{e}\Bigg (\frac {1}{T_e}+\frac {1}{T_i}\Bigg)
\frac {M'}{M} n_{d0}Z_d\nonumber \\
& D &=\eta_c  \Bigg (\frac {n_{e0}}{T_e}+\frac {n_{i0}}{T_i}\Bigg )
+n_{d0} \frac {|I_{e0}|}{e}\Bigg (\frac {1}{T_e}+\frac {1}{T_i}\Bigg ),
\label{delz1}
\end{eqnarray}
and the primes denote derivatives with respect to $z$ the
vertical direction. We then write down the longitudinal and
transverse components of the dust momentum equation (i.e. of
equation (\ref{gh})), as
\begin{mathletters}
\begin{eqnarray}
&&(1-i\omega \tau_m) \Big[(-i\omega +\nu) {\delta u_1}
+ {i k}\frac{\delta P}{Mn_{d0}} - \frac {Z_d e}{M} (ik \delta \phi)
 \Big]\nonumber\\
&=& -\frac {1}{Mn_{d0}}\eta_l  k^2 \delta u_1
\label {ghl}
\end{eqnarray}
\begin{equation}
(1-i\omega \tau_m) \Big[(-i\omega +\nu) {\delta u_3}
+\frac {\delta Z_d}{M}e {E_0} \Big]
= -\frac {1}{Mn_{d0}}\eta  k^2 {\delta u_3},
\label {ghs}
\end{equation}
\label{ghls}
\end{mathletters}
where, $\eta_l=\frac{4}{3}\eta+\zeta$.
In the limit $\omega\tau_m>>1$,  using equations (\ref{delz})--
(\ref{ghls}), we obtain the dispersion relation for the coupled
shear--compressional mode, as
\begin{eqnarray}
&&\Big [  \omega^2+i\omega \nu +i\omega \frac {eE_0}{MD} a_2
+\frac {eE_0}{MD} a_3 -C_{sh}^2k^2\Big]
\Big [ \omega^2+i\omega \nu -C_{DA}^2 k^2\Big]\nonumber\\
&-& i\omega k^2 \frac{eE_0}{MD} \frac {a_1 Z_d Z_d' n_{d0}}{MD}
+k^2\frac {eE_0}{MD}\frac {a_1 M'}{M}\big(C_d^2+C_{da}^2\big)
=0,
\label{disp}
\end{eqnarray}
\noindent where $\;$ $C_{sh}^2=(\eta /Mn_{d0}\tau_m)$,
$C_{da}^2=( Z_d^2 n_{d0} \eta_c / MD)$ and
$C_{DA}^2=C_d^2+C_{da}^2+ ( \eta_l / Mn_{d0}\tau_m)$. In the above
equation the expression in the first set of brackets represents
the dispersion relation for the transverse shear wave, the second
set of brackets contains the compressive mode dispersion relation
and the final two terms denote the coupling between the two
branches. We will now study the behaviour of the shear mode in
the presence of the charge inhomogeneity and the coupling to the
compressive mode.

\section{Shear Wave Instability}
In the limit when the coupling to the compressive wave is weak,
so that the last two terms in the dispersion relation (\ref{disp})
can be neglected, we can obtain the roots for the shear branch as,

\begin{equation}
\omega
 =-\frac {i}{2}\Big ( \nu + \frac{eE_0}{MD} a_2 \Big )
\pm \Big [ k^2 C_{sh}^2-\frac{eE_0}{MD}a_3-\frac {1}{4}
 \Big ( \nu + \frac{eE_0}{MD} a_2 \Big )^2\Big]^{1/2}.
\label{shearan}
\end{equation}
\noindent In the absence of the inhomogeneities and the collision
term, this is the basic shear wave described in \cite{kaw}. The
collisional term introduces wave damping. The inhomogeneous terms
introduce two important modifications. The term proportional to
the mass (size) inhomogeneity contributes to the real part of the
frequency whereas the charge inhomogeneity term can drive the
wave unstable if $E_0 a_2 <0$ (i.e., $E_0 Q_0'<0$) and the
threshold condition $\nu < |\frac {eE_0}{MD} a_2|$ is satisfied.
Physically, this instability arises because of delayed charging
effect, the same physical mechanism which was used by
Nunomura {\it et al} \cite{takamura} to explain the observed
instability of single particle vertical displacement in their
sheath experiments.  Specifically, the charge on the vertically
oscillating dust particle in the shear wave propagating in the
inhomogeneous plasma, is always different from the equilibrium
value $Z_d$ because of the finite charging time $\eta_c^{-1}$.
This perturbation is of order  $\delta Z_d \simeq Z_d'\delta
u_3/\eta_c$ and leads to an energy exchange between the shear
wave and the ambient electric field at a rate $\delta Z_d E_0
\delta u_3 ^{*} \approx |E_0 Z_d'||\delta u_3|^2/\eta_c$. When
this energy gain by the shear wave exceeds the loss rate due to
collisions $\approx \frac {\nu M}{2} |\delta u_3|^2$, we have an
instability. This gives us the approximate threshold condition
described above. If we express the dust neutral collision
frequency, $\nu$ in terms of the ambient neutral pressure as
$\nu=p\big(\frac{2 m_n}{T_n}\big)^{1/2}\frac{\pi a^2}{M}$, our
threshold condition is functionally identical to that derived by
Nunomura {\it et al}\cite{takamura} on the basis of physical
arguments. The only substantial difference is their use of
exponential charging time which follows from our equation
(\ref{fluct}) viz. $\delta Z_d
\approx (\delta u_3 Z_d'/\eta_c)[1-exp(-\eta_c t)]$; since we have
assumed the frequency of the shear mode $\omega <<\eta_c$, we use
the asymptotic condition described above.

We now demonstrate that for the collective shear mode being
described here, the coupling to the compressional dust acoustic
wave due to the last two terms in  equation (\ref{disp}) is very
crucial; thus the above single particle results are strongly
modified by the hydrodynamic treatment. A simple analytic result
clearly demonstrating the modification is obtained by neglecting
$\omega^2+i\omega\nu$ compared to $k^2C_{DA}^2$ in the second
bracket of equation (\ref{disp}); this is reasonable when the
wave--vector $k$ is not too small. In this limit, the shear modes
are described by the root
\begin{eqnarray}
\omega
& =&-\frac {i}{2}\Big ( \nu + \frac{eE_0}{MD}\Big( a_2
+a_1\frac{Z_d Z_d'}{MD} \frac {n_{d0}}{C_{DA}^2}\Big ) \Big )\nonumber \\
&\pm& \Big [{k^2 C_{sh}^2-\frac{eE_0}{MD}\Big (a_3
-a_1\frac {M'}{M}\frac {(C_d^2+C_{da}^2)}{C_{DA}^2}\Big)-\frac {1}{4}
 \Big ( \nu + \frac{eE_0}{MD}
\Big (a_2
+a_1\frac{Z_d Z_d'}{MD} \frac{n_{d0}}{C_{DA}^2}\Big )
 \Big )^2 \Big]^{1/2}}.
\label{shearcan}
\end{eqnarray}
We thus note that the threshold condition and the growth rates
are significantly modified by the inclusion of coupling to
compressional waves. In order to quantitatively illustrate the
effect of coupling terms, we now present a detailed numerical
investigation of the dispersion relation equation (\ref{disp}).
It is generally the case that the bulk viscosity coefficient
$\zeta$ is negligible compared to the shear viscosity coefficient,
$\eta$, particularly in the one component plasma (OCP)
limit\cite{ich} and so we shall drop it in our calculations.
Further, the viscoelastic relaxation time, $\tau_m$, is given as
\cite{kaw},
\begin{equation}
\tau_m=\frac {4 \eta}{3 n_{d0} T_d (1-\gamma_d \mu_d +
\frac{4}{15} u)}
 \end{equation}
with $\gamma_d$ as the adiabatic index and the compressibility,
$\mu_d$ defined through (\ref{eos}). We assume the gradient of
the equilibriated dust charge to be of the form,
$Z_d'={Z_d}/{L_{Z}}$, the mass gradient to be of the form
$M'=M/L_{M}$ where $L_{Z} \sim L_{M} = L$ is a few Debye lengths. In our
computations, we choose $L \approx 5$ times the Debye length, which is the
typical order of magnitude as observed experimentally
\cite{nunomura}. For further computations, we introduce the
dimensionless quantities,
\begin{eqnarray}
\hat \omega=\omega/\omega_{pd};\;\;\;
\hat \nu=\nu/\omega_{pd};\;\;\;
\hat k=kd;\;\;\;
\hat \tau_m=\tau_m \omega_{pd};\;\;\;
\nonumber \\
\hat \eta=\frac {\eta}{Mn_{d0} \omega_{pd} d^2};\;\;\;
{\hat C}_\alpha^2=C_\alpha^2/( \omega_{pd}^2 d^2);
\alpha \equiv sh,d,da,DA,\;\;\;
\nonumber \\
e_0=\frac {eE_0}{MD} \frac {a_2}{\omega_{pd}};\;\;\;
e_1=\frac {eE_0}{MD} \frac {a_3}{\omega_{pd}^2};\;\;\nonumber \\
e_{01}=\frac {a_1 Z_d Z_d' n_{d0}}{a_2 MD}
 \frac {1}{\omega_{pd}^2 d^2};\;\;
 e_{11}=\frac {eE_0}{MD}a_1 \frac{M'}{M}(\hat C_d^2
 +\hat C_{da}^2)\frac {1}{\omega_{pd}^2},
\end{eqnarray}
\noindent where $\omega_{pd}$ and $d$ are the dust plasma
frequency and the inter--grain distance respectively. \noindent
The dispersion relation for the shear mode (\ref {disp}) can then
be written as
\begin{eqnarray}
&&\Big [ \hat \omega^2+i\hat \omega \hat \nu +i\hat \omega e_0+e_1
 -\hat C_{sh}^2 \hat k^2\Big]
\Big [ \hat \omega^2+i\hat \omega \hat \nu -\hat C_{DA}^2 \hat k^2\Big]
\nonumber\\
&-& i\hat \omega \hat k^2 e_0 e_{01}+\hat k^2 e_{11}=0,
\label{disps}
\end{eqnarray}
Equation (\ref{disps}) has been solved numerically for the shear
mode roots and some typical results are presented in figures (1)
and (2). Figures (1a) and (1b) display a comparison of the
dispersion curve for the shear mode ($\hat \omega_R$ {\it vs}
$\hat k$ and $\hat \gamma$ {\it vs}  $\hat k$ for fixed values
of $e_0 = -0.0008$ and $e_{1}=-0.05$), with and without the
inclusion of the coupling to the compressional mode. The various
fixed parameter values corresponding to these curves are $\hat
C_{sh}^2=0.02$, $\hat C_{DA}^2=0.4$, $\hat \nu =0.0004$ and
$e_{01} = 0.3$, $e_{11}= -0.01$ when the coupling is on. The
choice of these numerical values for the dimensionless parameters
$\hat \nu$, $\hat k$, $e_0$, $e_{01}$, $\hat C_{sh}$ and $\hat
C_{DA}$ has been guided by the magnitude of these quantities
observed in some of the laboratory plasmas\cite{chu,nunomura}. It
is seen from these plots that there is a substantial influence of
the compressional mode coupling, described through the parameter,
$e_{01}, e_{11}$, on the growth rate and the real frequency of
the shear wave emphasizing the importance of the collective
physics of coupling to the compressional mode. We next plot in
figure (2) the gas pressure, $p$, versus $n_{e0}$ profiles for
various values of $\gamma$, the imaginary part of $\omega$.
Plotting the $\gamma=0$ curve, we get a threshold relation
between $p$ and $n_{e0}$, where we fix the other parameters as
follows -- dust radius, $r_d$=2.5 microns, the inter--grain
distance, $d$=430 microns, $T_e = T_i \simeq 1eV$, $kd =1$, and
dust mass density, $\rho_d$=2.5 gms/cm$^3$. We see that the
qualitative trend of the curve is similar to that observed in the
single particle instability studies of \cite{takamura}
illustrating the commonality of the underlying physical
mechanism. However it should be emphasized that the experiment in
\cite{takamura} did not observe any collective excitations and
their equilibrium consisted of a monolayer of equal sized
particles. The equilibria of \cite{chu,nunomura} are more
appropriate for observing collective excitations of shear waves
and our theoretical results can be usefully employed in such a
situation. In Fig.(2) we have once again highlighted the
significance of the coupling to the compressive wave, in this
case for its effect on the threshold values, by displaying the
uncoupled threshold and growth rate curves (dashed curves). Note
that the influence of the coupling is to raise the threshold
value at low values of $n_{e0}$ (i.e. a higher value of $p$ is
needed to excite the instability) whereas it reduces the
threshold at the higher end of the $n_{e0}$ scale. The rest of
the curves displayed in the figure (2) correspond to the various
positive values of $\gamma$, which correspond to the situation
where the shear mode is excited and saturates at some values.
These figures are again qualitatively similar to the curves
obtained in \cite{takamura} for various saturation amplitudes.
However a direct comparison is again not appropriate for the
reason discussed above and also because our calculations are
linear and cannot provide any quantitative results about
nonlinearly saturated amplitudes.

\section{Conclusion and Discussion}
To summarize, in this paper we have investigated the stability of
a low frequency shear mode in an inhomogeneous dusty plasma in
the strongly correlated regime. The equilibrium dust cloud has
both an inhomogeneity in the dust charge distribution and in the
dust mass distribution (arising from a distribution in the sizes
of the dust particles). The shear mode in such a plasma undergoes
two significant modifications. Its real frequency is shifted by a
contribution from the mass inhomogeneity and the dust charge
inhomogeneity can drive it unstable through the dynamics of dust
charge fluctuations in a manner very similar to the instability
of the vertical motion of single particles in a plasma sheath as
observed in the recent experiment of Nunomura {\it et
al}\cite{takamura}. The finite charging time, $\eta_c^{-1}$ of the
dust particles plays a critical role in the instability. We also
show how collective effects due to coupling with the
compressional modes strongly modify the threshold conditions for
the instability as well as its growth rate and real frequency.
Our calculations have been carried out in the hydrodynamic
formalism including viscoelastic effects and we have neglected
any kinetic effects. Our results are therefore strictly valid in
the low frequency limit. Finite corrections arising from kinetic
effects can occur at higher frequencies and wave numbers. This
has recently been demonstrated for the compressive dust acoustic
mode in a dusty plasma from a kinetic calculation based on the
dynamic local field correction (DLFC) method\cite{murillo}. Such
corrections, if any, for the transverse shear mode has not yet
been done and needs to be examined. \\

Finally we would like to remark that the transverse dust shear
mode which is a collective mode of the strongly coupled plasma
regime has only been observed in computer simulations till now;
its detailed experimental investigation is therefore of great
current interest. Such waves can be excited in inhomogeneous dust
clouds that have been obtained in the experiments carried out with
varying grain sizes \cite{chu,nunomura}. It would be of interest
therefore to look for the wave features discussed in our model
calculations in controlled propagation experiments on such
equilibria.  It is also apparent that free energy sources, such
as ion beams, which may readily couple with the compressional
waves may also be useful for exciting the more interesting shear
waves in the strongly coupled inhomogeneous plasma. Investigation
of these and related effects are in progress.

\vfil
\eject

\newpage
\begin{figure}
\psfig{file=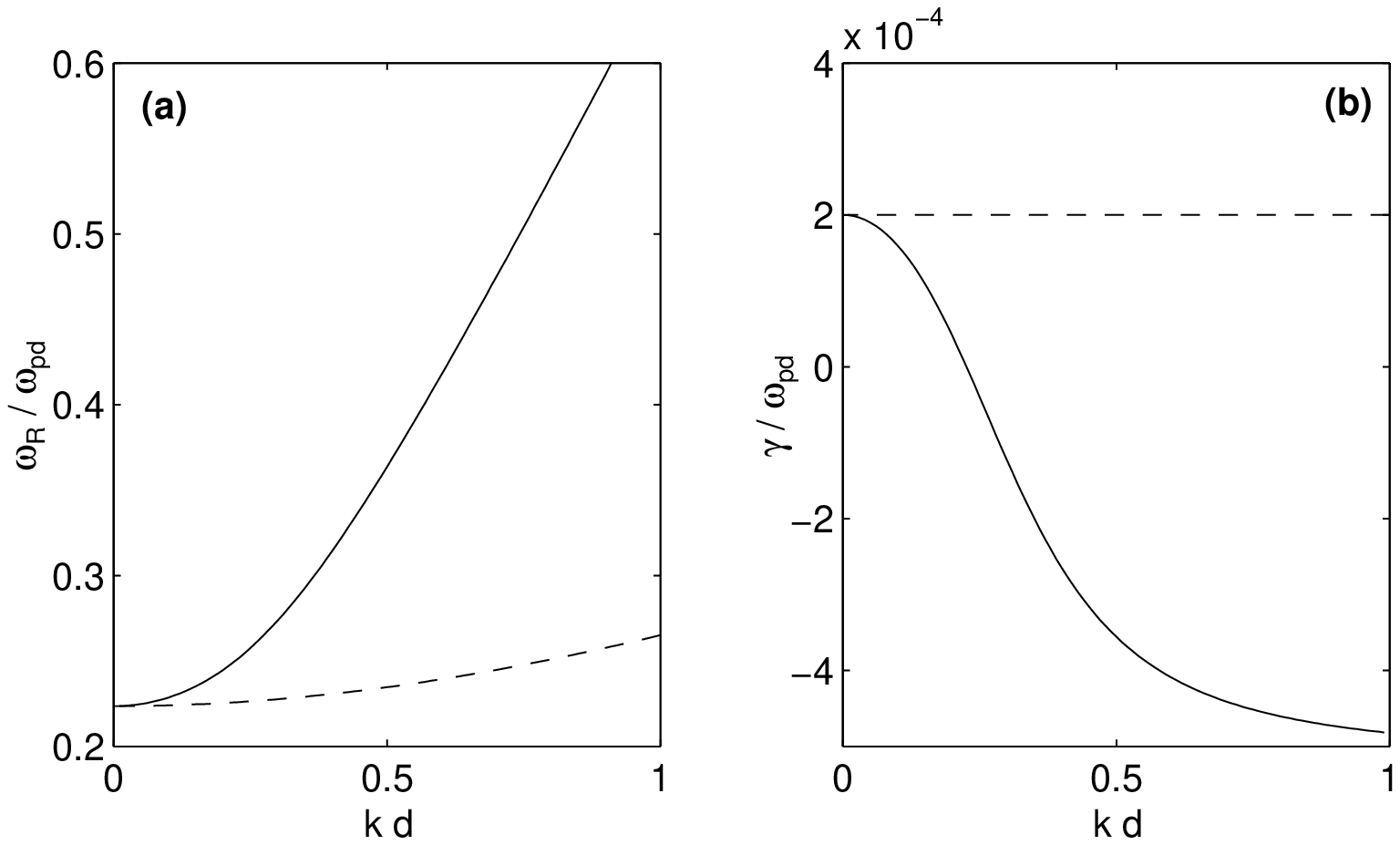,width=16cm,height=12.5cm}
\caption{(a) The normalized real frequency
and (b) the normalized imaginary frequency,  ${\it vs.}$ the
normalized wave number for the shear mode with $e_0=-0.0008 $,
$e_{01}= 0.3$, $e_{1}= -0.05$, $e_{11}=-0.01$
(solid curves). The dashed curves are for $e_{01} = e_{11} =0$ and
correspond to the uncoupled shear mode.}
\end{figure}

\begin{figure}[h]
\psfig{file=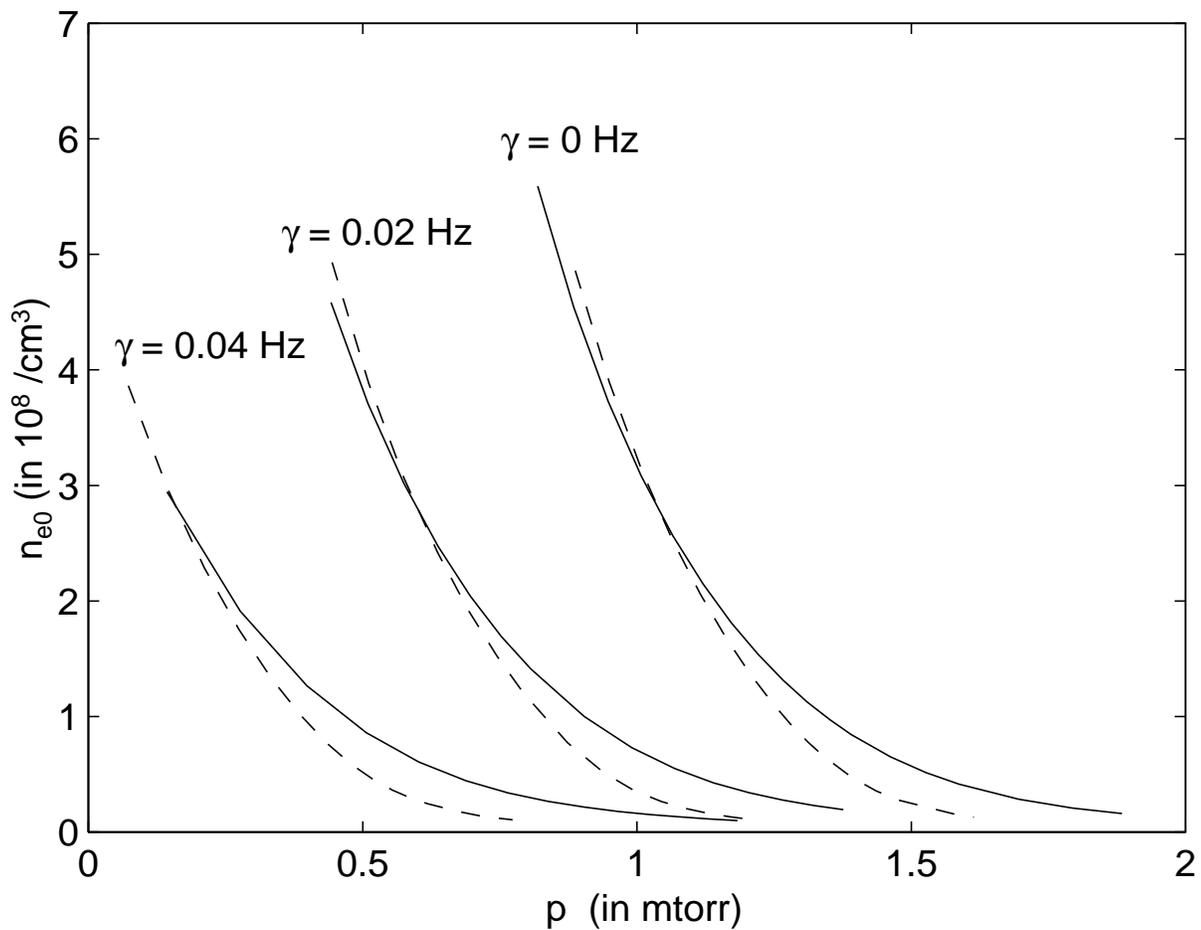,width=16cm,height=12.5cm}
\caption  {The electron number density $n_{e0}$
(in units of $10^8/{\rm cm}^3$) is plotted as a function of the
gas pressure, $p$ (in mtorr) for various values of $\gamma$. For
comparison, the accompanying dashed curves display the situation
when the coupling to the compressional mode is neglected.}
\end{figure}
\end{document}